\documentclass[useAMS,usenatbib]{mn2e}

\usepackage{graphicx}
\usepackage{appendix}
\usepackage{psfrag}
\usepackage{amsmath,amssymb} 
\usepackage{threeparttable}
\usepackage{float}
\usepackage{subfigure}
\usepackage{afterpage}
\numberwithin{equation}{section}

\title[ALFALFA H{\sc i} Data Stacking III]{ALFALFA H{\sc i} Data Stacking 
III. Comparison of environmental trends in  {\hi} gas mass fraction and specific star formation rate}  
\author[S. Fabello]{Silvia Fabello$^{1}$\thanks{fabello.silvia@gmail.com}, Guinevere Kauffmann$^{1}$\thanks{gamk@mpa-garching.mpg.de}, Barbara
  Catinella$^{1}$, Cheng Li$^{2}$, \newauthor  Riccardo Giovanelli$^{3}$, 
 Martha P. Haynes$^{3}$
 \\
$^{1}$Max-Planck Institut f\"{u}r Astrophysik, D-85741 Garching, Germany\\
$^{2}$Max-Planck-Institute Partner Group, Shanghai Astronomical
Observatory, Nandan Road 80, Shanghai 200030, China\\
$^{2}$Centre for Radiophysics and Space Research, Cornell University, Ithaca, NY 14853, USA\\
}

\begin{document}          

\newcommand{\hi}{H{\sc i}}
\newcommand{\Mhi}{M$_{\rm HI}$}
\newcommand{\Mst}{M$_\star$}
\newcommand{\must}{$\mu_\star$}
\newcommand{\col}{NUV$-r$}
\newcommand{\cix}{$C$}
\newcommand{\vel}{km$\,$s$^{-1}$}
\newcommand{\df}{D$_n$(4000)}
\maketitle

\label{firstpage}
\begin{abstract}
It is well known that both the star formation rate  and the cold gas content
of a galaxy depend on the local density out to distances 
of a few Megaparsecs.
In this paper, we  {\em compare} the environmental density dependence of the    
atomic gas mass fractions of nearby galaxies with the density dependence  of  their           
central and global  specific star formation rates. 
We stack {\hi} line spectra extracted from the Arecibo Legacy Fast ALFA survey    
centered on  galaxies with UV imaging from GALEX  and optical imaging/spectroscopy from  SDSS.
We use these stacked spectra  to evaluate the mean atomic gas mass fraction of 
galaxies in bins of stellar mass and local density. 
For galaxies with stellar masses less than $10^{10.5} M_{\odot}$, 
the decline in  mean atomic gas mass fraction 
with  density is stronger than the decline in   
mean global and  central specific star formation rate.
The same conclusion does not hold for more massive galaxies.
We interpret our results as evidence for ram-pressure stripping of
atomic gas from the outer disks of low mass satellite galaxies. 
We  compare our results with the semi-analytic recipes of
Guo et al. (2011) implemented on the Millennium II simulation. 
These models assume that 
only the diffuse gas surrounding satellite galaxies is  stripped, 
a process that is often termed ``strangulation". We show that these
models predict relative trends in atomic gas
and star formation that are in disagreement with observations.  
We use mock catalogues generated from the simulation to predict the halo masses of
the {\hi}-deficient galaxies in our sample. We conclude that ram-pressure stripping
is likely to become effective in dark matter halos with masses greater
than $10^{13}$ M$_{\odot}$.

\end{abstract}


\section*{Introduction}

Systematic studies of the dependence of galaxy properties on
environment began with  analyses of the 
relation between galaxy morphology and local density  
(Oemler 1974; Dressler 1980). It later became evident that star formation is
more strongly affected by environment
than morphology (Hashimoto et al. 1998). 
Large surveys, for example
the Sloan Digital Sky Survey (SDSS; York et al. 2000)  and the
Two-degree Field Galaxy Redshift Survey (2dFGRS; Colless et al. 2001)
provided samples large enough to study the effects of local galaxy
density  on a multiplicity of galaxy properties. It was shown that the
star formation and structural  properties of galaxies  depend strongly
on their  mass (e.g. Kauffmann et al. 2003; Shen et al. 2003; Baldry
et al. 2004). Because of this, and because galaxy mass is itself
correlated with environment, it is important to bin galaxies by mass 
before studying how their properties vary with local density.                       
Several analyses showed that at fixed mass, the 
dependence of  colour/star formation on density is 
stronger than that of  structural properties such as concentration
index and stellar surface mass density (Kauffmann et al. 2004; Li et
al. 2006a;  
Bamford et al. 2009; Skibba et al. 2009). 
This suggests that
environmentally-driven processes lead to cessation of star formation
in a galaxy,  but do not strongly affect its structure. 

An alternative way of quantifying environmental effects is to study
how the properties of galaxies vary as a function of distance from 
the centers of groups and clusters. The ``center'' is usually defined
as the position of the brightest galaxy in the system and the
distance is usually scaled to the virial radius.
One result that has emerged from such studies is that 
at fixed clustercentric distance, galaxy properties do not  depend strongly 
on the mass of the group (Balogh et al. 2004; Van den Bosch et al. 2008).
This does not agree with a
scenario in which the ram-pressure effects are  responsible
for removing the cold interstellar medium in galaxies and shutting down
star formation. 
Ram-pressure depends on the square of the velocity at which
a galaxy is moving through the surrounding gas, so
it should operate more efficiently in galaxies in more massive dark matter halos.

In 1980, Larson, Tinsley \& Caldwell  suggested that the gas {\em
  envelopes} surrounding disks are most easily stripped by ram-pressure
effects when they  fall into a cluster. After a few Gyr, the galaxy
will have converted its available cold gas into stars and 
because there is  no  infall of new gas, the galaxy
will ``starve''  and stop forming stars. This 
mechanism has long been part of  semi-analytic models  of
galaxy formation (e.g. Kauffmann et al. 1993).
In recent years, the predictions of these   models have been compared with
data from SDSS and it has been found that the fraction of red
satellites is {\em higher} in the models than in the data
(Kimm et al. 2009; Weinmann et al. 2010). 
Star formation quenching timescales must therefore be quite long (2-3 Gyr)  in satellite
galaxies (Wang et al. 2007). In the optical astronomy community,
``slow gas starvation'' has thus come to
be accepted as the main physical process determining how galaxies
evolve in dense environments. 

We note, however, that optical  studies  present a biased picture of
how environment affects galaxies. In disk galaxies, 
atomic gas generally extends to substantially
larger radii than the stars. Ram-pressure  will act primarily  on
low density atomic gas in the outer regions of galaxies, rather than the dense
molecular gas in their central regions.   
The first systematic studies of the dependence of the
atomic gas content of galaxies on environment   
(Haynes, Giovanelli \& Chincarini 1984; Giovanelli \& Haynes 1985;
Gavazzi 1987) found that  
disk galaxies in clusters exhibit a deficiency in {\hi} content
that strongly increases towards the cluster centre
and that  matches the predictions of the 
the ram-pressure stripping model introduced by
Gunn \& Gott (1972). Integrated CO observations of cluster galaxies
suggested that there is no deficiency of molecular gas in
{\hi}-deficient galaxies (Kenney \& Young 1989), but resolved studies
showed some evidence of CO depletion when the {\hi} is stripped to
within the optical disk (Vollmer et al. 2008; Fumagalli et
al. 2009). Both  results support the idea that ram-pressure stripping
primarily affects gas in the outer regions of galaxies, proceeding
inward. 
Subsequent work by Gavazzi (1989), Cayatte et al. (1990), Kenney, Van
Gorkom \& Vollmer (2004), Chung et al. (2009) emphasized the frequent
presence of cluster galaxies with  truncated {\hi} disks which,
together with examples of disturbed {\hi} morphologies and one-side
tails, also support the mechanism of ram-pressure.  
Finally, SPH simulations (Abadi, Moore \& Bower 1999; Vollmer 2009) of galaxies
orbiting through the intra-cluster medium then  demonstrated 
that ram-pressure can in fact be responsible for
distorted {\hi} disks similar to those seen in the
observations. 

The degree to which ram-pressure may affect the interstellar medium of
galaxies outside the rich cluster environment is not yet well
understood. Available samples have generally been too small to
quantify environmental effects across a large dynamic range in  local
density or dark matter halo mass. 
The state-of-the-art blind {\hi} survey, the Arecibo Legacy Fast ALFA
survey (ALFALFA; Giovanelli et al. 2005),  does detect {\hi} in
galaxies in environments spanning a range of environments from     
voids to rich clusters, but ALFALFA is still a shallow survey 
so it will only detect gas-rich galaxies at redshifts greater than $\sim 0.02$. 
In this paper, we employ the stacking technique described in
Fabello et al. (2011a, hereafter Paper I) to
study the \emph{average} cold gas content of galaxies as a function
of local density for a sample of $\sim 5000$ galaxies with redshifts
in the range $0.025 < z < 0.05$. By comparing the variation of
the {\hi} gas fraction with environment with the variation of their 
total and  central specific star formation rates,
we aim to constrain the environments in which ram-pressure stripping
effects become important.  
As we have discussed, ram-pressure stripping will
preferentially affect the low density outer disks of galaxies,
which are dominated by atomic gas where star formation is inefficient
(Bigiel et al. 2010). 

We begin by describing the data and the density estimator that we use 
in this analysis. 
We then compare the \emph{relative} decrease in {\hi} gas fraction and
specific star formation rate as a function of local density  
and  compare our results with the results of  semi-analytic models
implemented on  the Millennium II Simulation (Boylan-Kolchin et al. 2009). Discussion and conclusions
are presented in the final section.

\section{The Sample}\label{sample}
Our galaxies are selected from the  ``parent sample'' of the GALEX Arecibo
SDSS Survey (Catinella et al. 2010), which is a volume-limited
sample of $\sim$12000 galaxies selected from the SDSS
main spectroscopic sample  with stellar masses greater than
 $10^{10} M_{\odot}$ and redshifts in the range 0.025$< 
z <$0.05, and which  
lie in the intersection of the footprints of the SDSS Data Release 6
(Adelman-McCarthy et al. 2008),
the GALEX Medium Imaging Survey (Martin et al. 2005)
and the ALFALFA survey. 
We make use of  \textit{sample A} defined in Paper I, which 
consists of 4726 galaxies in the ALFALFA
40\% dataset (Haynes et al. 2011).  
Only 23\% of \textit{sample A} targets  are  detected
by ALFALFA. We employ a stacking  
technique, which allows us to include the many non-detections. 
We refer the reader to Paper I for a 
comprehensive description of the stacking method; in this paper,
we only provide a brief summary.
Before proceeding, we also describe the parameters
used in this analysis; these include stellar mass, global and fibre  specific star
formation rates ($\S$\ref{ssfr_par}), and our  adopted environmental tracer
($\S$\ref{s:tracer}).

\subsection{ALFALFA data stacking}\label{stack}
ALFALFA is a blind {\hi} survey that used the ALFA multibeam receiver
at the Arecibo telescope to scan 7000 deg$^2$ of the sky
over the velocity interval v[km$\,$s$^{-1}]\simeq$[-2500; 18000]
(i.e. out to z$\sim$0.06). The data acquired are stored as smaller
three dimensional cubes of dimension 2.4$^{\circ}\times$2.4$^{\circ}$
on the sky and 5500 km$\,$s$^{-1}$ in velocity ``depth''. The stacking
process that we apply to ALFALFA data includes a series of steps, which can be 
summarized as follows:\\

{\bf a. Create a catalogue of {\hi} spectra}\\
All our targets are selected from the SDSS spectroscopic survey, so we
know their position on the sky and their redshift. We select the
ALFALFA data-cube which contains the target and integrate the signal
from the galaxy over a sky region of 4$'\times$4$'$ (our targets
are always smaller than the telescope beam, whose FWHM is
$\sim$3.5$'$). For each spectrum we measure the root mean square
(\textit{rms}) noise, which is used later as a weight.

{\bf b. Stack spectra}\\
We  co-add the signals from N different sources located at
different redshifts. First, we shift each spectrum to the target galaxy rest
frequency, so each spectrum is centered at zero velocity. We stack
together the spectra $S_i$ (i=1,..N) using their $w_i=$1/\textit{rms}$^2$ as a
weight, so that the final spectrum $S_{stack}$ is: 
\begin{equation}
S_{stack} = \frac{\Sigma^N_{i=0}S_i \cdot w_i}{\Sigma^N_{i=0} w_i}.
\end{equation}
If we recover a signal in the stacked spectrum, we measure the
integrated emission between the two edges of the {\hi} profile, which
are defined manually for each spectrum. If there is no detection, we
evaluate an upper limit, assuming a 5$\sigma$ signal with a width of 300
km$\,$s$^{-1}$, smoothing the spectrum to 150
km$\,$s$^{-1}$. 

{\bf c. Evaluate {\hi} gas fractions}\\
Our aim is to compute the average {\hi} content of a given sample
of galaxies, so we are interested in converting our recovered signal
into an {\hi} mass and subsequently  into an average {\hi} gas fraction. Once
we measure an {\hi} flux, we  estimate the corresponding {\hi} mass using: 
\begin{equation}
\frac{\mathrm{M_{HI}}}{\mathrm{M_{\odot}}}=\frac{2.356\times10^5}{1+z}\left(\frac{D_L(z)}{\mathrm{Mpc}}\right)^2\left(\frac{S_{int}}{\mathrm{Jy~km~s^{-1}}}\right)  
\end{equation}
where $D_L(z)$ is the luminosity distance and $S_{int}$ the integrated
{\hi} flux. The {\hi} gas fraction is simply defined as
{\Mhi}/{\Mst}.  
Note that we  weight each spectrum
before stacking by {\Mst}$^{-1}(1+z)^{-1}D_L(z)^2$, to convert
it into a measure of ``gas fraction'' (see also the discussion in Paper
I, Appendix A). \\

At the median redshift of \textit{sample A}, the size of the
Arecibo telescope beam corresponds to physical scales of 0.15 Mpc and
may include more galaxies than the targeted  one. For our gas fractions
to be reliable, we apply a correction for contamination from close
companions, as described in detail in Appendix A. 
In summary, for each galaxy which lies inside a region of
the beam size $\pm\,300\,${\vel} around the main target, we estimate
its gas fraction from photometry using  the relation between colour,
stellar mass surface density and {\hi}  gas fraction found by
Zhang et al. (2009), and subtract it from the
{\hi} mass of the target galaxy. The corrections are
always  smaller than a few percent  even  in the highest density bins.

\subsection{Galaxy Parameters}\label{ssfr_par}
The optical parameters we use are drawn  from the MPA-JHU SDSS DR7
release of spectrum measurements or from Structured Query Language
(SQL) queries to the SDSS DR7 database
server\footnote{See http://www.mpa-garching.mpg.de/SDSS/DR7/ and
  http://cas.sdss.org./dr7/en/tools/search/sql.asp}. 
We use UV/optical colours derived from convolving the SDSS images
to the same resolution as the GALEX images; this is  described in 
more detail in Wang et al. (2010). 

The parameters  used in this paper are the following.

\begin{description}
\item[{\bf Stellar masses {\Mst}}] are derived from SDSS
photometry using the spectral energy distribution (SED) fitting
technique described in Salim et al. (2007)  with a Chabrier (2003) initial
mass function.  

\item[{\bf Specific star formation rates (sSFR)}] are defined as the
star formation rate per unit stellar mass, SFR/{\Mst}
[yr$^{-1}$]. We use two different measures of sSFR to trace
different regions of the galaxies: a fibre and a global specific
star  formation rate. 1) The {\em fibre specific star formation
rate} is measured inside the 3-arcsecond SDSS fiber, and is
therefore characteristic of the inner regions.
At the median redshift of the sample ($z=0.035$), the 3 arcsecond
diameter fiber subtends a physical length scale of 2.1 kpc. The radius
enclosing 50\% of the $r$-band light for galaxies with stellar masses between
$10^{10}-10^{11}$ M$_{\odot}$ ranges from 2 to 3 kpc.  So the fraction of the total galaxy
light going down the fibre will be around 5-10\%.
We acquire the fibre specific star formation rates
from the MPA-JHU SDSS DR7 release of spectrum 
measurements\footnote{http://www.mpa-garching.mpg.de/SDSS/DR7/}. Briefly,
they are evaluated from the spectrum emission lines, to which a
grid of photo-ionization models from \citet{Charlot01} is fitted, following the
methods described in \citet{Brinchmann04}. For objects whose signal is contaminated by AGN
emission or for objects with low S/N  emission lines,
the  SFRs are derived indirectly from the 4000 {\AA}  break strength.
2) the {\em global specific star  formation rate} 
is obtained by applying a spectral energy  distribution (SED)
fitting technique to the five optical and two GALEX  UV total
flux measurements.  
A more thorough discussion of this procedure is
presented in Saintonge et al. (2011).
\end{description}

\subsection{ Local density estimator}\label{s:tracer}
\begin{figure}
\centering
\begin {tabular}{cc}
    \includegraphics[height=4.2cm]{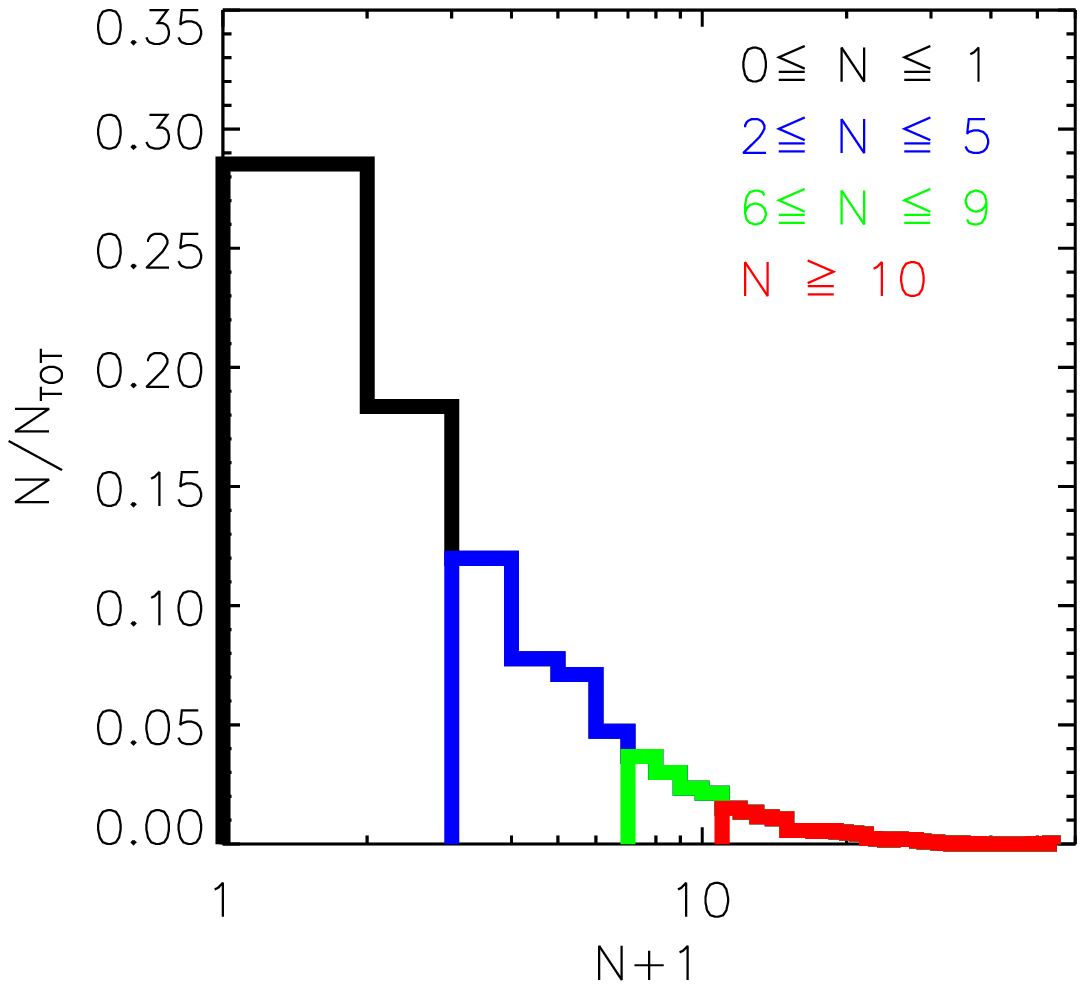}&
    \includegraphics[height=4.2cm]{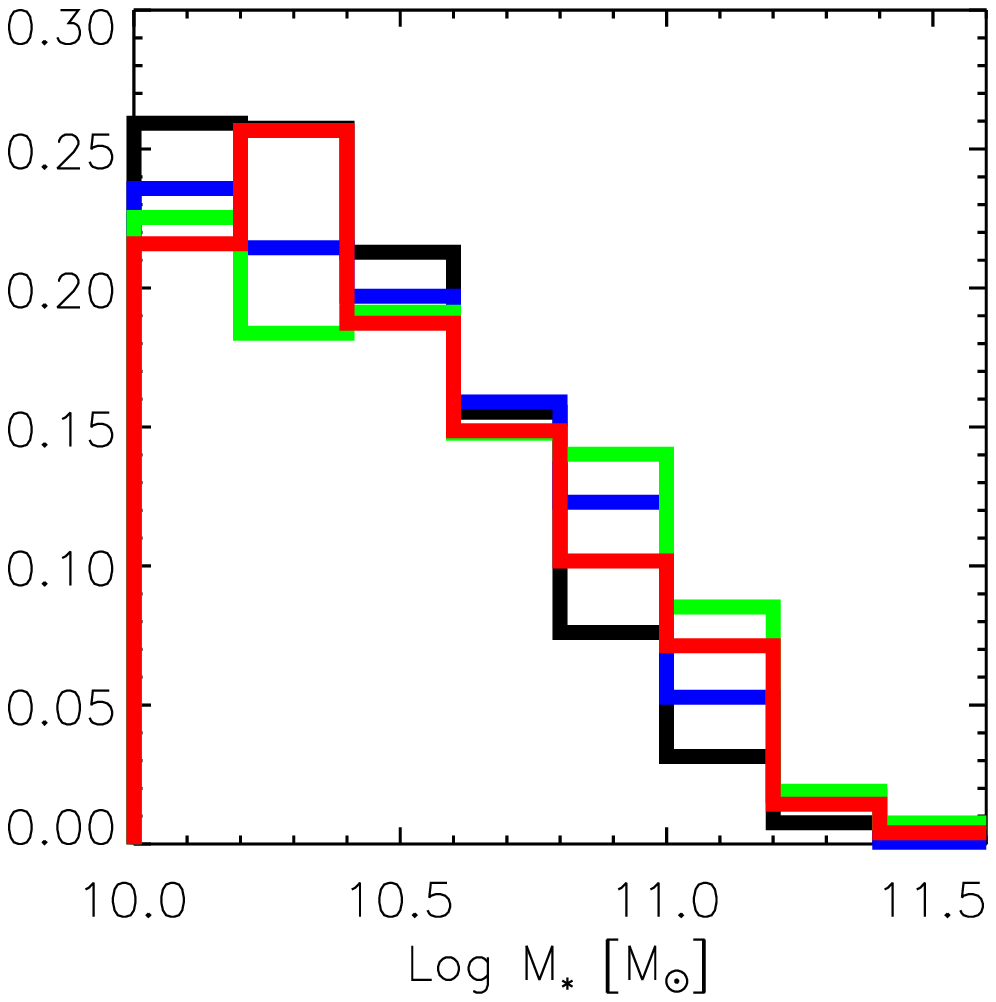}\\
  \end{tabular}\caption{
    \textit{Left:} normalized distribution of the
    density parameter $N$ (plotted as $N+1$ for convenience). The colours
    indicate the four density bins we will use throughout the
    paper, as reported in the legend. \textit{Right:} Normalized
    stellar mass distribution for galaxies in  each of the density bins.}
    \label{en:fig:histo}
\end{figure}
Similar to other past studies of galaxy environment
(e.g. Kauffmann et al. 2004; Blanton \& Berlin 2007; Thomas et al. 2010), we
define a density parameter for each galaxy as the number of neighbours
with Log {\Mst}[M$_{\odot}$]$\geq 9.5$ located inside a ``cylindrical'' 
aperture of 1 Mpc radius and $\pm\,500\,${\vel} depth, centred on the
target. Environmental effects are strongest if density is evaluated
on scales comparable to the typical virial radii of the haloes
hosting the galaxies in the sample.    1 Mpc   
is somewhat larger than the typical virial
radius of haloes hosting L$_*$ galaxies, but on  
smaller scales the number of tracer galaxies becomes too  small and
Poisson noise dominates. 

We search for neighbours with Log {\Mst}[M$_{\odot}$]$\geq 9.5$
in the MPA-JHU DR7 spectroscopic 
catalogue.  We note   that  two SDSS fibres cannot
be closer than 55 arcseconds, so we might miss close companions. In order
to correct for this ``fibre collision'' effect, we follow 
the  approach of Li et al. (2006b), who  measured the angular
two-point correlation function for the SDSS spectroscopic sample
$[w_z(\theta)]$ and for the parent photometric sample
$[w_p(\theta)]$. The ratio:
 
\begin{equation*} F(\theta) =
  \frac{1+w_p(\theta)}{1+w_z(\theta)} \end{equation*}
is used to
correct for the effect of fibre collisions. We adopt the 
correlation functions from Li et al. (2006a)  and weight  each neighbour
by $F(\theta)$, where $\theta$ is the angular separation from the main
target. In practice,  fibre collision corrections on the
  estimate of N are very small 
on average ($\sim3$\%). 

The distribution of the density
parameters $N$ derived for \textit{sample A} galaxies 
is shown in Figure \ref{en:fig:histo}, left panel (note that we actually plot
$N$+1 for convenience). Just under 50 \% of the  galaxies in our
sample have zero or one neighbour. The right panel 
shows the normalised stellar mass distribution for
galaxies with 0-1, 2-5, 6-9 and $>10$ neighbours, following the colour
legend in the left panel. As can be seen,
the stellar mass distributions do not change very much between the different density bins.
As we will show, this is because most galaxies in the bin with $>10$ neighbours
are not in rich cluster environments, but in dark matter halos of
moderate mass ($\sim 10^{13}$ M$_{\odot}$).

\begin{figure} 
  \centering
  \begin {tabular}{cc}
    \includegraphics[width=4.cm]{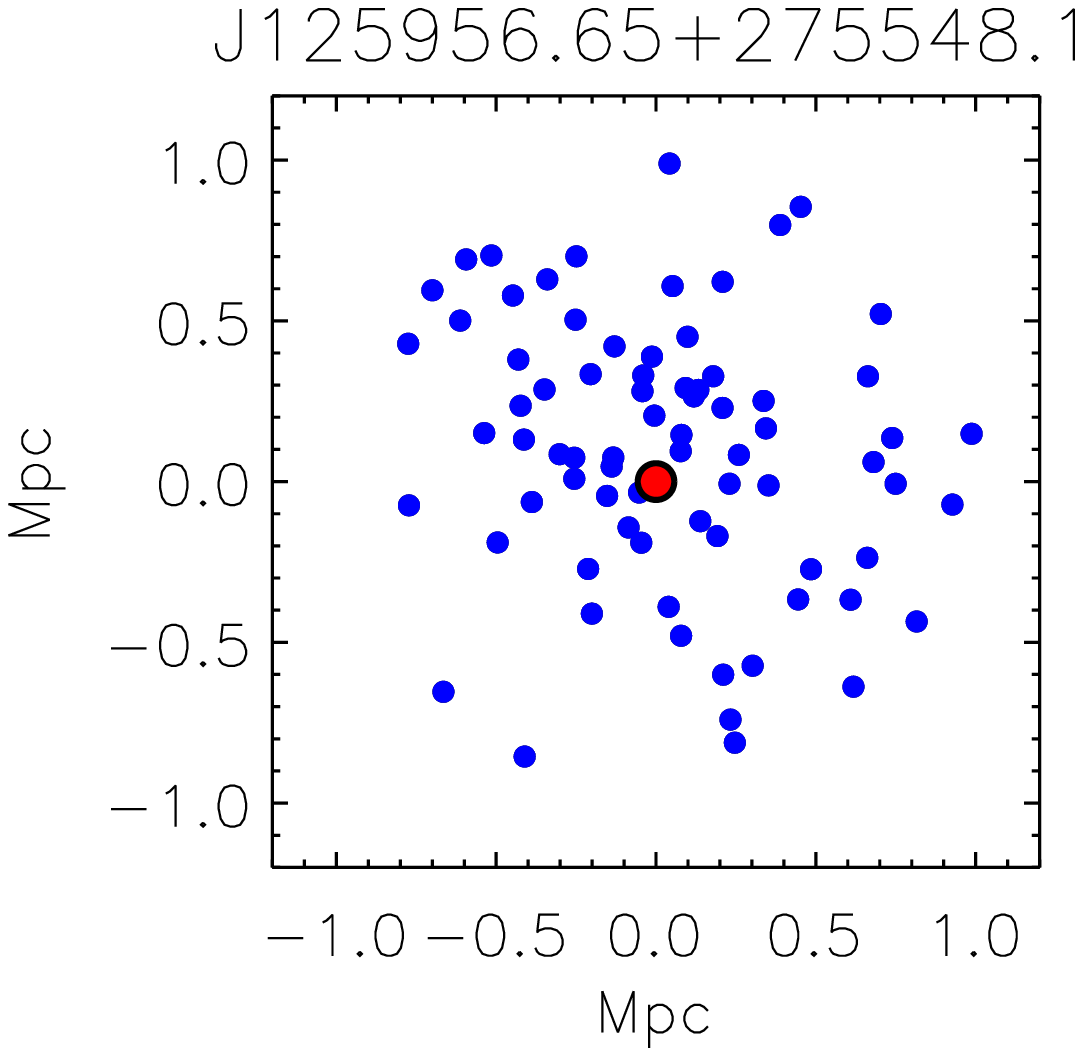}&
    \includegraphics[width=4.cm]{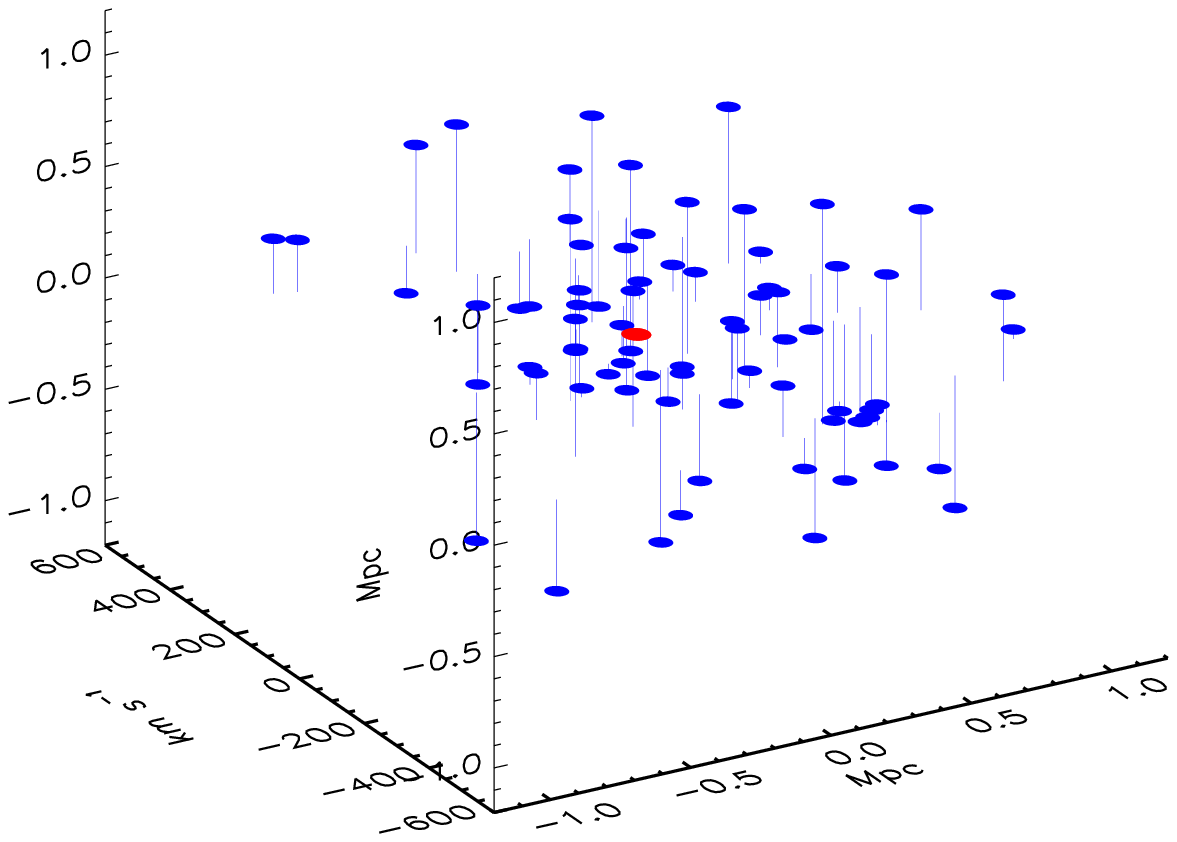}\\
    \includegraphics[width=4.cm]{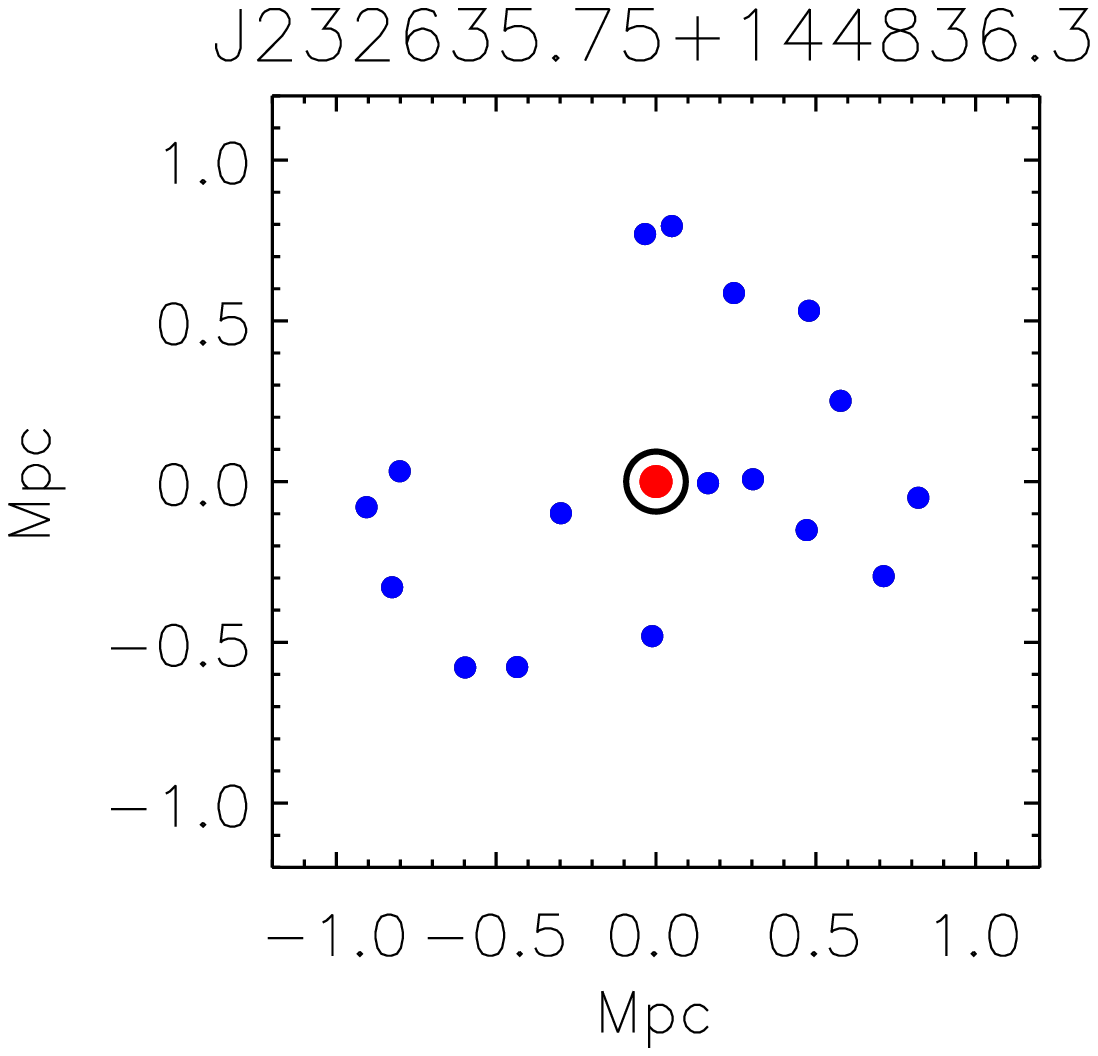}&
    \includegraphics[width=4.cm]{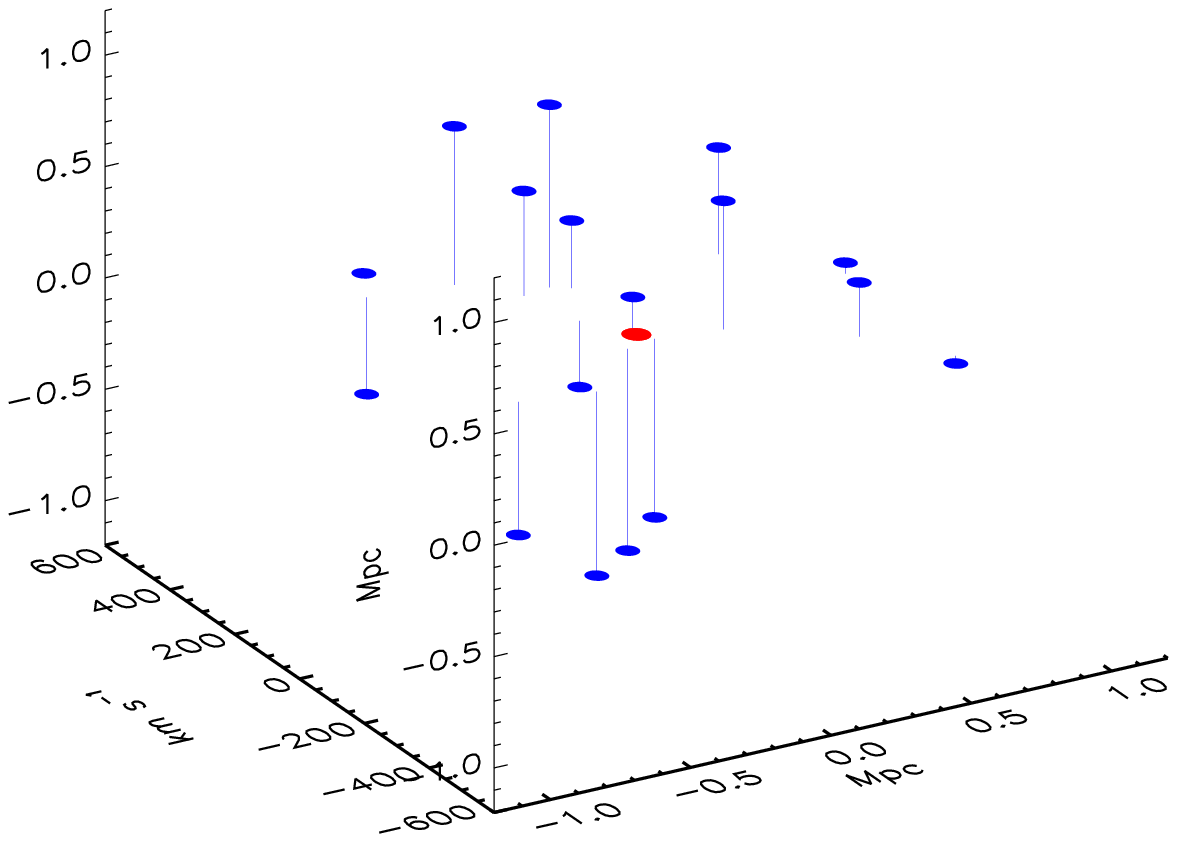}\\
  \end{tabular}\caption[Sky distribution of two groups 
    in our sample.]{Sky distribution of neighbours around two galaxies    
    in our sample. \textit{Left:} sky projection as a function of
    distance (in Mpc) from the central target. The size of the
    Arecibo beam is overplotted as a black circles on the central galaxies
    (red dot, SDSS name on top). 
    \textit{Right:} three-dimensional
    view of the same group, with the redshift/velocity component
    added. The top row shows the galaxy in the richest environment in our sample, the
    bottom row a typical galaxy in the density bin with $N>10$ .}\label{en:fig:fig01}
\end{figure}

In Figure 2, we show the spatial distribution of
galaxies found around two objects in our sample with $N>10$.   
On the left, we plot the galaxies projected  on the
sky in units of Mpc. The target galaxies are represented
by the red dots and the black circles around them indicate the
physical size of the Arecibo beam at the redshift of the target. 
On the right, we show the three-dimensional distributions of the neighbours. 
The galaxy in the  top row is the one with largest number of
neighbours in \textit{sample A} ($N=72$). 
This galaxy  lies in the far 
outskirts of the Coma cluster (the mean redshift of the Coma cluster is 
$cz\,=\,$6853$\,\pm\,$1082$\,${\vel} (Colless \&
Dunn 1996), and this galaxy lies at  $cz\,=\,$7860$\,${\vel}). 
Most of the Coma cluster is actually outside our redshift
range, however. In the bottom row
of Figure \ref{en:fig:fig01}, we show a typical  galaxy with
$N=17$. As can be seen, such a galaxy is not in a cluster, but in a group. 

\section{Comparison of the density dependence of {\hi} mass 
fractions and global/fibre sSFRs} 
\begin{figure*}
  \centering
  \begin{tabular}{c}
    \includegraphics[width=16cm]{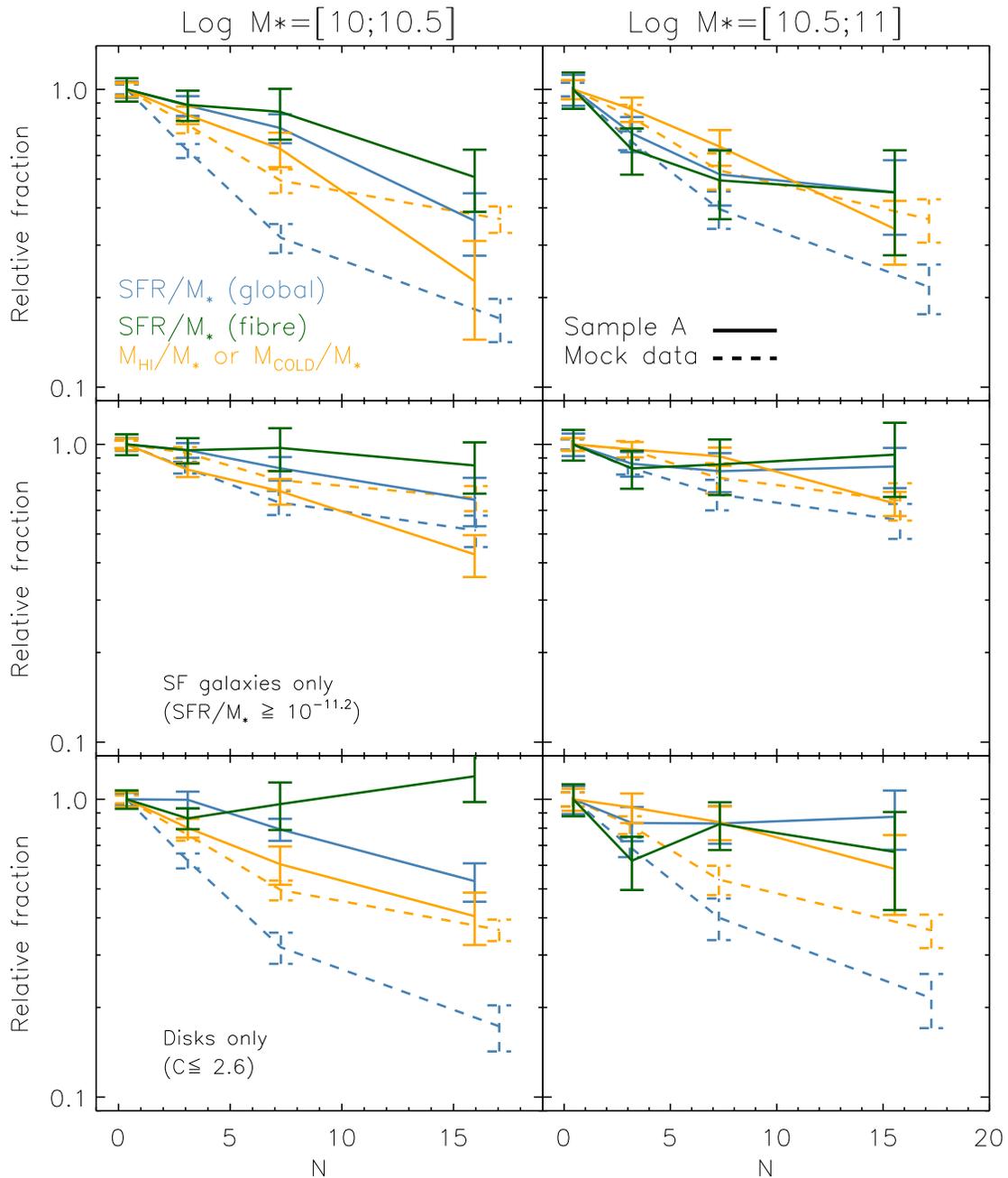}
  \end{tabular}\caption[Environmental effects and different properties
  of galaxy]{Comparison of the relative dependence of the
{\hi} gas fraction (orange), the
global specific star formation rate (blue) and the fibre specific star
formation rate  (green)  as a function of local density for 2
bins of {\Mst}, as labelled at the top of the diagram.  \textit{First row}:
Results are shown for all
galaxies. \textit{Second row}: Results are shown for star-forming galaxies 
with Log SFR/{\Mst}$\geq-11.2$. \textit{Third row}: Results are shown for disk-dominated
galaxies with $C\leq2.6$. Solid lines show results obtained
by stacking \textit{sample A} galaxies. Error bars are evaluated by bootstrap re-sampling
the galaxies in the stack. Dashed lines show the results derived from semi-analytic model  
mock catalogues (see Section 2.1).}\label{en:fig:compd} 
\end{figure*}

It is well known that both  star formation rates and {\hi} gas fractions
are smaller in galaxies in dense environments (e.g. Balogh et
  al. 2004; Cortese et al. 2011).  
As we have discussed, a  {\em comparison} of how the {\hi} mass
fractions and the specific star formation rates of galaxies  measured
in their centers and in their outer regions depend on environment
should constrain the physical origin of these effects. 
Because the atomic gas extends to larger radii and lower densities
than the gas that traces  the young stars, a comparison of
the local density dependence of  atomic gas mass fractions and central
sSFRs  should also provide considerable insight.  
In a pure starvation scenario, where  there is no replenishment of the cold gas
in the disk from cooling of the hot halo, we would expect the {\hi} 
and the star formation  to decrease at the same rate as a function of
density. If ram-pressure stripping of the atomic  gas is important, we
would expect the {\hi} to exhibit a  stronger  environmental dependence. 

In this section, we compare the {\em relative} decrease in {\hi} mass fraction,
global and central specific star formation rates as a function of local
density for galaxies in different stellar mass bins. 
We divide our sample (we do not
  consider galaxies with {\Mst}$>10^{11}$ M$_{\odot}$ because of limited
  statistics) into two bins of stellar mass  and four bins of local density. For each bin, we compute
M$_{\mathrm{HI}}$/{\Mst}, global and fibre specific star formation
rates. We  then scale these values by dividing by the value
measured for the lowest density bin ($0\leq N\leq1$)  at the same stellar mass.
In this way, we compare 
the relative  decrease of each quantity with density. 
 
Results are shown in Figure 3. The orange solid lines represent the
{\hi} gas mass fractions, the blue solid lines show 
the global specific star formation rates, and the green solid lines  the
fibre specific star formation rates. Errorbars are computed by
bootstrap resampling 80\% of the galaxies included in the stacks. The dashed lines
show results from models and  are discussed in detail in the next section.  

The top panels show results for the whole \textit{sample A},
while the second and third panels investigate different
 sub-populations to gain better insight into the processes at work. In
particular, the
middle row focuses on galaxies with Log
SFR/{\Mst}$\,$[yr$^{-1}$]$\,\geq-11.2$.
The cut is chosen because it is the minmum of the bimodal distribution
of specific star formation rates of galaxies in our sample 
(see  von der Linden et al. 2010). The bottom panels
show results for disk-dominated 
galaxies, selected from \textit{sample A} to have concentration
indices less than 2.6. The concentration index is
defined as the 
ratio of the radii enclosing 90\% and 50\% of the $r$-band light and
is quite tightly correlated with the  bulge-to-total ratio B/T of the galaxy (Gadotti
  2009). $C<2.6$ corresponds to B/T$<0.3$ (see also Weinmann et al. 2009).
As can be seen, for  galaxies with stellar masses in the range
$10 <$Log {\Mst}$< 10.5$, there is a clear ordering in that the {\hi} mass
fraction declines most steeply as a function of local density, followed
by the global and fibre specific star formation rate.  
For galaxies with stellar masses in the range $10.5 <$Log {\Mst}$< 11$,
there is no similar  ordering.  

\begin{figure*}
  \centering
  \begin{tabular}{c}
    \includegraphics[width=16cm]{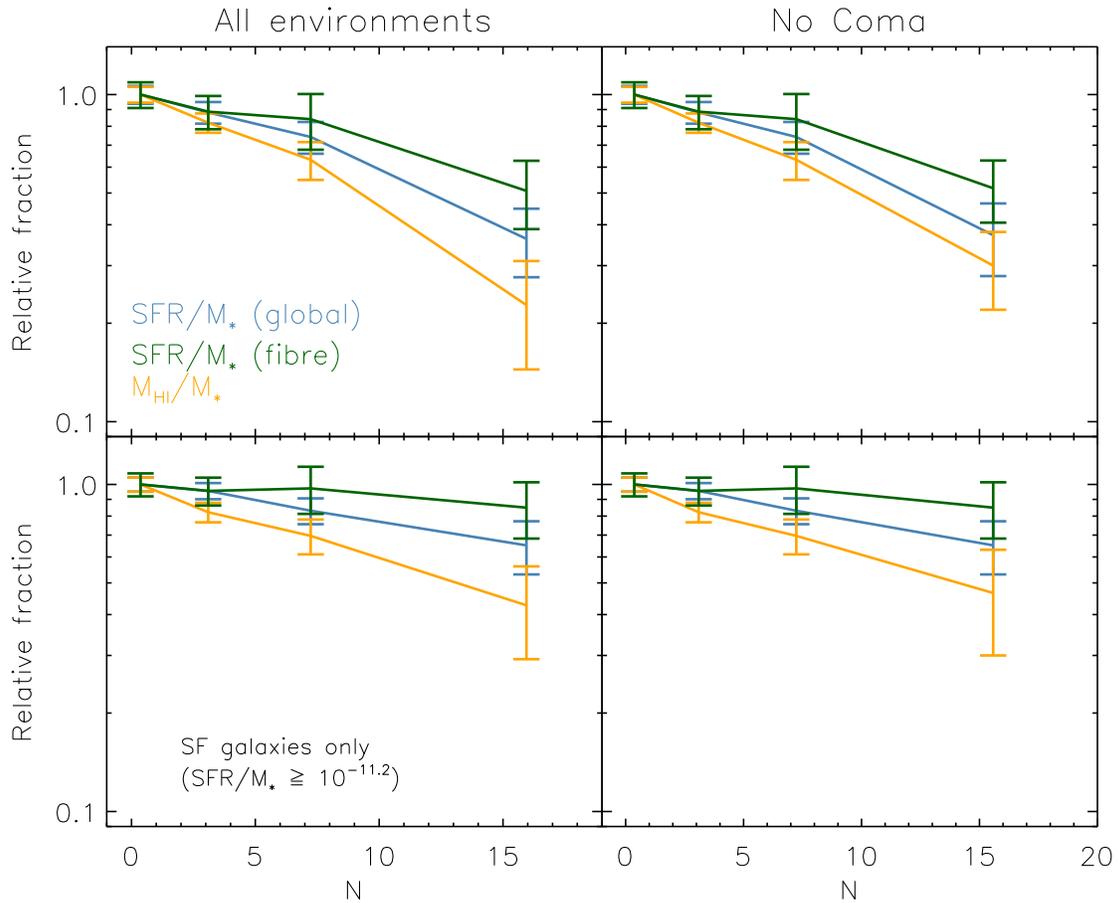}\\
\end{tabular}\caption[Effect of removing Coma galaxies]                
{The relative dependence of
{\hi} gas fraction (orange), the
global specific star formation rate (blue) and the fibre specific star
formation rate  (green)  as a function of local density for  galaxies
with $10 <$Log {\Mst}$< 10.5$. The left panels show results for
galaxies in \textit {sample A}, all (top) or only the star forming
ones (bottom); the right panels
show analogous results when galaxies in the vicinity of the Coma cluster are removed.}   
\label{en:fig:compd} 
\end{figure*}

As we have discussed, \textit{sample A} does include galaxies located at the
edge of the Coma cluster. In order to check the extent to which
the effects seen in  
Figure 3 are caused by a sub-population of cluster galaxies,
we exclude Coma cluster galaxies from our analysis and replot the
mean gas fraction and specific star formation rate curves in 
Figure 4. In practice, we  exclude 12 galaxies with
stellar masses greater than $10^{10}$ M$_{\odot}$  within 3 degrees
and $\pm 2000\,${\vel} of the center of the Coma cluster.   
As can be seen, the decrease in both the average {\hi} gas mass fraction
and the global and specific star formation rates as a function of 
density becomes somewhat weaker when the Coma galaxies are discarded,
but the ordering remains the same.
As we will discuss later, our results thus support a scenario in which ram-pressure stripping 
affects low mass galaxies in {\em moderate-density environments}. 

As can be seen by comparing the results for star-forming galaxies
and  disk-dominated
galaxies with the results obtained for ``all'' galaxies, a significant
part of the decrease in {\hi} gas mass fraction as a function of 
density is driven by processes acting on star-forming, disk-dominated systems.
In contrast, in the high mass bin, there is no significant
decline in {\hi} mass fraction as a function of local density for
star-forming, disk-dominated galaxies. Most of the decline in {\hi} mass
fraction for massive galaxies seen in the top-right panel   
must thus be driven by an increase
in the fraction of passive, early-type galaxies in denser environments.  

It is also interesting to compare the decrease in the {\hi} gas mass
with the decrease in the central specific star formation rate 
measured within the fiber.  
The decrease in central sSFR is driven by
passive, early-type galaxies in both stellar mass bins. The processes
acting on atomic gas disks apparently do not affect the central sSFRs at all,
at least  over the
range of local densities probed by our sample.

\subsection{Comparison with models}\label{comp:mock}

So far, we have tentatively interpreted our results as possible evidence for
the effect of ram-pressure  acting on the atomic gas in  disks in low
mass galaxies 
in environments characteristic of galaxy groups. We do not, however,
have information about the spatial distribution of the gas in our galaxies.
We have made an ``ansatz''  that on average the gas will be more spatially
extended than the star formation and that ram-pressure will more
strongly affect the {\hi} gas in the outskirts of galaxies than the
star-forming (i.e. molecular) gas. 
 
In this section, we compare our data with results from semi-analytic models
where ram-pressure stripping of the cold interstellar medium of galaxies
is {\em not taken into account}.  We show that these models predict
relative trends in {\bf cold} gas mass fractions and specific star formation rates
that do not agree with the observations.   
 
We make use of outputs from the semi-analytic models of 
Guo et al. (2011, hereafter Guo11)  implemented on the Millennium II
simulation,
which are publically available for download 
at http://www.mpa-garching.mpg.de/galform/millennium-II/. 

In these models, the cold interstellar medium  of a galaxy is distributed in a
disk with size that scales as the product of the virial radius and
the spin parameter of its host halo.  Cold gas is  
supplied both by infall of diffuse gas and by gas 
from accreted satellites. Cold gas is depleted by star formation and
reheated to the hot phase by supernovae.
The total star formation rate in the disk scales with
its total cold gas content following a simplified version 
of the Kennicutt (1998) law. Stars will not form in a disk unless the total
cold gas mass exceeds a certain critical value, which is set by the condition that 
its surface density is large enough for the gas to be gravitationally  
unstable (Toomre 1964). In the model, this stability criterion is a
global rather than a local one.  

Tidal effects, ram-pressure stripping and radio AGN feedback act  on the   
\emph{diffuse gas} associated with each galaxy and prevent it from cooling,
condensing and forming new stars.  
As shown in Figure 3 of Guo11, the implementation of these quenching processes
in the models lead to trends in the fraction of actively
star forming galaxies as a function of projected distance 
from the centres of rich clusters that are in relatively good agreement with
observations. 

We make use of  a set of 100 mock SDSS galaxy catalogues from the Guo11 model that match   
both the sky mask and the magnitude and redshift limits of the SDSS DR7 sample. 
These are the same mock catalogues that were used for
interpreting SDSS data in a recent paper by Li et al. (2012). 
Detailed description of the methodology for constructing the
mocks can be found in Li et al. (2006b; 2007). 

We begin by extracting a volume from each mock catalogue that is exactly matched
in redshift and sky area coverage to ALFALFA \textit{sample A}.
We select galaxies with  Log {\Mst}$\,$[M$_{\odot}]=[10;11.5]$, and then
apply the same method used for the observations to compute a local density
parameter ($N$) by counting the neighbours more massive than
{\Mst}$=10^{9.5}\,$M$_{\odot}$ inside a  cylinder of 1$\,$Mpc radius
and depth $\pm\,500\,${\vel}.   

In Figure 5, we compare the fractions of
quenched objects, defined as galaxies 
with SFR/{\Mst}$\le
10^{-11.2}\,$yr$^{-1}$, in the real sample and in the mock
catalogues. Results are shown for two stellar mass bins and we
normalize the quenched   
fraction N$_q$/N in each density bin to the value for isolated objects
with $N=0$. Red lines show results for the real data
and black dotted lines are for the model galaxies. 
The  error bars on the data are obtained by boot-strap
resampling. For the models, the error bars represent
the variance over the 100 mock catalogues. 
As can be seen, the models and the data agree quite well.
The fraction of quenched low mass galaxies is somewhat
higher in the data than in the models in the highest density bin  
($N>10$), where the disagreement is at about 2$\sigma$ level.

\begin{figure}
\centering
\includegraphics[width=8.5cm]{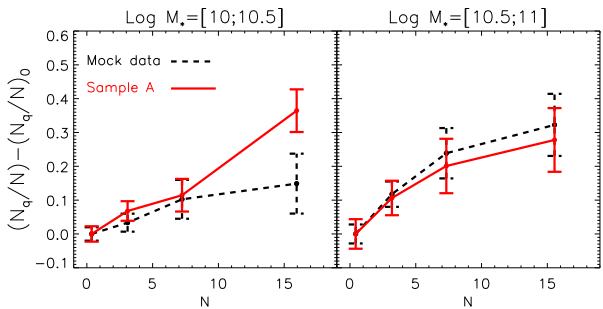}
\caption[Comparison of mock and real quenched
fractions]{Comparison of the density dependence of the fraction of
  ``quenched'' galaxies with SFR/{\Mst}$\le 10^{-11.2}\,$yr$^{-1}$ in
  the data (solid red lines) and in the mock catalogues (dashed black). 
We plot the fraction of quenched objects with respect to that
found for isolated objects:  (N$_q$/N)-(N$_q$/N)$_0$. 
The two different panels show results for two  different
stellar mass bins, as reported on top.}\label{en:fig:quench} 
\end{figure} 

As we have discussed, in the models there are a variety of processes that
quench star formation in galaxies. Feedback from radio AGN acts to prevent gas
from cooling onto the {\em central galaxies} of dark matter halos, whereas tidal
effects and ram-pressure stripping of the hot gas will affect ongoing
star formation in {\em satellite galaxies}.  

In Figure 6, black curves show the fraction of model galaxies
of given stellar mass that are
satellites as a function of the density parameter $N$.
As can be seen, in the lower stellar mass bin ($10 <$Log {\Mst}$< 10.5$),
the fraction of satellite galaxies increases very strongly as a function of $N$.
For $N=5$, 60 percent of galaxies are satellites and for $N>10$,
the fraction of satellites is around 0.9. This means that the
analysis of the full sample of low mass galaxies in this paper
is likely to  probe the physical
processes relevant to satellite rather than to central galaxies.
In the higher stellar mass bin ($10.5 <$Log {\Mst}$< 11$),
the fraction of satellites at intermediate values of $N$ is smaller.
However, for $N>10$, the fraction of satellite systems still reaches
values greater than 0.8.

Green and blue curves show satellite fractions
as a function of local density for disk-dominated and star-forming 
galaxies, respectively. The satellite fractions for disk-dominated 
galaxies are almost the same as for the whole sample. However, for
star-forming galaxies,  satellite fractions are much lower.
In fact, in the models, the star-forming galaxy samples are dominated
by central galaxies at all densities, so such samples may not be efficient
probes of tidal and ram-pressure stripping processes.

\begin{figure}
\centering
\includegraphics[width=8.5cm]{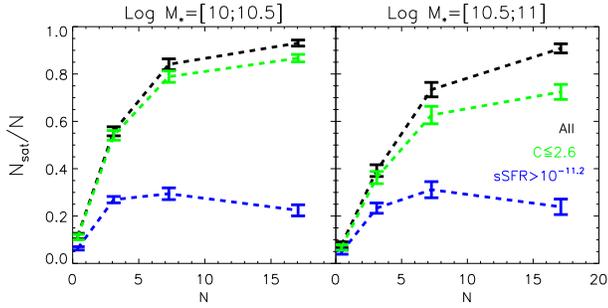}
\caption[Fraction of satellite galaxies as a function of
density]{The mock catalogues generated from the
semi-analytic models are used to calculate the 
fraction of galaxies of given stellar mass that are
satellite systems as  a function of
density parameter $N$. Black curves show results for
all galaxies in the mass bin, green curves show results for
disk-dominated galaxies, and blue curves show results for
star-forming galaxies. 
The error bars indicate the variance in in the satellite
fraction estimated from the 100 mock catalogues.}                                 
\label{en:fig:sats} 
\end{figure} 

We now study  relative trends in cold gas mass fraction and specific star formation rate
in the models, and compare these to what is seen in the data. 
In Figure 3, orange and blue dashed lines
show the cold gas fractions and global specific star formation rates  
of the model galaxies as a function of density parameter $N$.
As can be seen, the main result is that the specific star formation
rate decreases more strongly as a function of local density than the cold gas
mass fraction.  This is seen in both stellar mass bins. 
The effect is very strong in the top and bottom panels, which show results for all galaxies
and for disk galaxies selected according to their bulge-to-disk ratios. 
The effect becomes considerably weaker if the sample is
restricted to galaxies with ongoing star formation (middle panel), because
these are mainly central galaxies
\footnote {We note that the models do not include molecular gas as a
separate phase. The average molecular-to-atomic ratio in present
day galaxies is about a third (Saintonge et al. 2011). If the atomic gas
fraction is observed to drop by a factor to 0.5 of its field value and the
molecular gas is unaffected, the total cold gas mass fraction will drop
two-thirds of its field value. Accounting for the molecular component
will not, however, cause a reversal in the trend between cold gas fraction   
and specific star formation rate.}.

In the models, ram-pressure acts only
on the diffuse gas halo surrounding satellites.  One might think that this would imply  
that star formation and the cold gas remain closely coupled.  
The apparently puzzling result that the star formation (dashed blue
line) is more strongly affected by environment than the cold gas 
(dashed orange) is  a consequence of the fact that in  
galaxies where the cold gas mass has fallen below the threshold value,
star formation shuts down and cold gas is no longer consumed.

In Figure 7, we compare the relations between gas mass fraction versus
global specific star formation rate for model galaxies (black dotted) and \textit{sample A} galaxies (red
  solid) found in rich environments ($N\geq7$) where satellite
  galaxies dominate. We remind the reader that  \textit{sample A} gas 
fractions are estimated using stacked spectra and  errors
are computed via bootstrapping. For the models, we plot the  
mean {\hi}  gas mass fraction and the errors represent the variance between the 100
  mock catalogues. 
As can be seen, the relation between gas fraction and specific star formation
rate is much steeper in the real data than in the models (the red
triangle indicates an upper limit). In the real Universe,
satellite galaxies are not usually left with an inert reservoir of low-density cold gas unable to
form stars. We suggest that this is because such reservoirs are more easily stripped
from the galaxy than currently assumed in the models.  

\begin{figure}
\centering
\includegraphics[width=8.5cm]{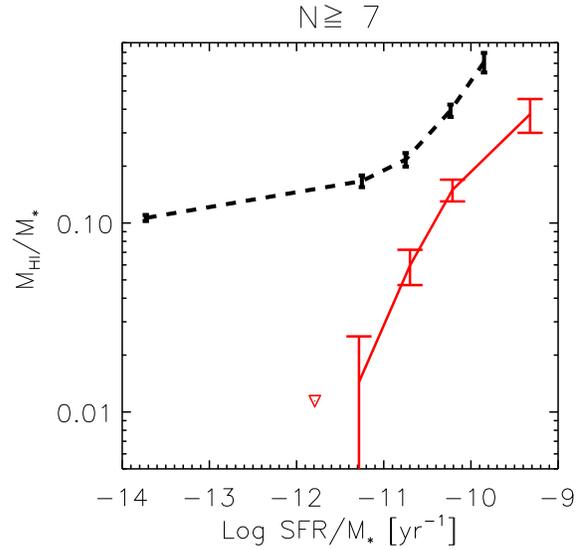}
\caption[Relation between gas mass fraction and specific star formation
rate]{The relation between {\hi}/cold gas mass fraction and global
specific star formation rate for model galaxies (dashed black) and for
galaxies in  
{\it sample A} (solid red) located in environments with $N\geq 7$.}
\label{en:fig:sfrmock} 
\end{figure} 

In summary, the implementation of star formation quenching processes
in the semi-analytic models 
produces relative trends in gas mass fraction and specific star
formation rate that disagree  
with observations. 

\section{Summary and discussion}\label{concl}

In this work, we have used a complete, volume-limited sample of
nearby galaxies with {\Mst}$>10^{10}$M$_{\odot}$, with 
coverage by the ALFALFA, SDSS and GALEX surveys, to study
how the average {\hi} content and the global and
central specific star formation rates of galaxies depend
on local density at fixed stellar mass. 

Our main new  result is that {\hi} gas mass fraction and
specific star formation rate do not scale with local density
in the same way.   
For galaxies with stellar masses less than $10^{10.5} M_{\odot}$ 
the atomic gas mass fractions decline most strongly as a function of density,
followed by their global and  central specific star formation rates.
The same ordering is not seen  for more massive galaxies.

In order to interpret this result, we compare our results with
mock galaxy catalogues generated using the semi-analytic recipes          
of Guo et al. (2011) implemented on high resolution cosmological simulations of structure
formation in a $\Lambda$CDM Universe. 
We demonstrate that the local density parameter that we have defined
is tightly correlated with the fraction of galaxies that are
{\em satellite} rather than central galaxies. 

In the Guo11 models, star formation in
satellite galaxies shuts down as a result of gas ``starvation'' --
tidal and ram-pressure forces remove the gaseous halos  surrounding
the satellites, and as a result, cooling and infall of new  gas onto these systems
ceases. The star formation rates in  satellites  decline as their
cold gas reservoir is used up. Eventually the surface density of cold gas falls
below the critical threshold value for star formation to occur, and
star formation stops entirely. The models thus predict that the
average specific star formation rates of galaxies should decrease with $N$ more strongly
than their {\hi} gas mass fractions, which is exactly the opposite to what
is seen in  observations. 

We suggest that the assumption
in the models that ram-pressure acts {\em only} on the diffuse gas surrounding
galaxies is wrong. A question one might ask is whether the
{\hi}-deficient galaxies in our 
sample are located in environments where ram-pressure stripping could
plausibly occur.   
In Figure 8,  we plot the distribution of
dark matter halo masses that 
host galaxies in our mock catalogue with stellar
masses in the range $10 <$Log {\Mst}$< 10.5$ and local environment
parameter $N\geq 7$ (solid curve) and $N<7$ (dotted curve). $N\geq 7$ corresponds to environments where the 
mean gas mass fraction of galaxies of this mass bin has dropped by more than a factor of two 
with respected to ``isolated'' galaxies with $N=0$.
As can be seen, the  majority of such galaxies are in dark matter  
halos with masses in the range $10^{13} - 10^{14} M_{\odot}$, i.e they are in
galaxy {\em groups} rather than clusters.
In contrast, most galaxies in the same stellar mass range with $N<7$ are located in dark matter
halos with masses less than $10^{12}$ M$_{\odot}$, which are not expected to have a hot gas
atmosphere (Birnboim \& Dekel 2003).
We conclude, therefore, that in order to bring the models into agreement
with observations, ram-pressure effects would need to strip atomic gas from galaxies in
dark matter halos more massive than $10^{13}$ M$_{\odot}$. 

\begin{figure}
\centering
\includegraphics[width=8.5cm]{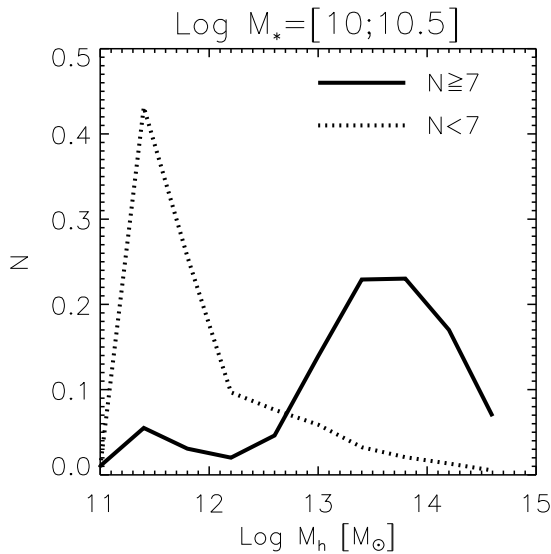}
\caption[Halo masses of sample A galaxies]{The distribution of dark
  matter halo masses that host galaxies in our mock catalogue with stellar
masses in the range $10 < \log M_* < 10.5$ and local environment parameter $N>7$ (solid
curve) and $N<7$ (dotted curve).}  
\label{en:fig:halom} 
\end{figure} 

\begin{figure} 
  \centering
  \begin {tabular}{c}
    \includegraphics[width=8.5cm]{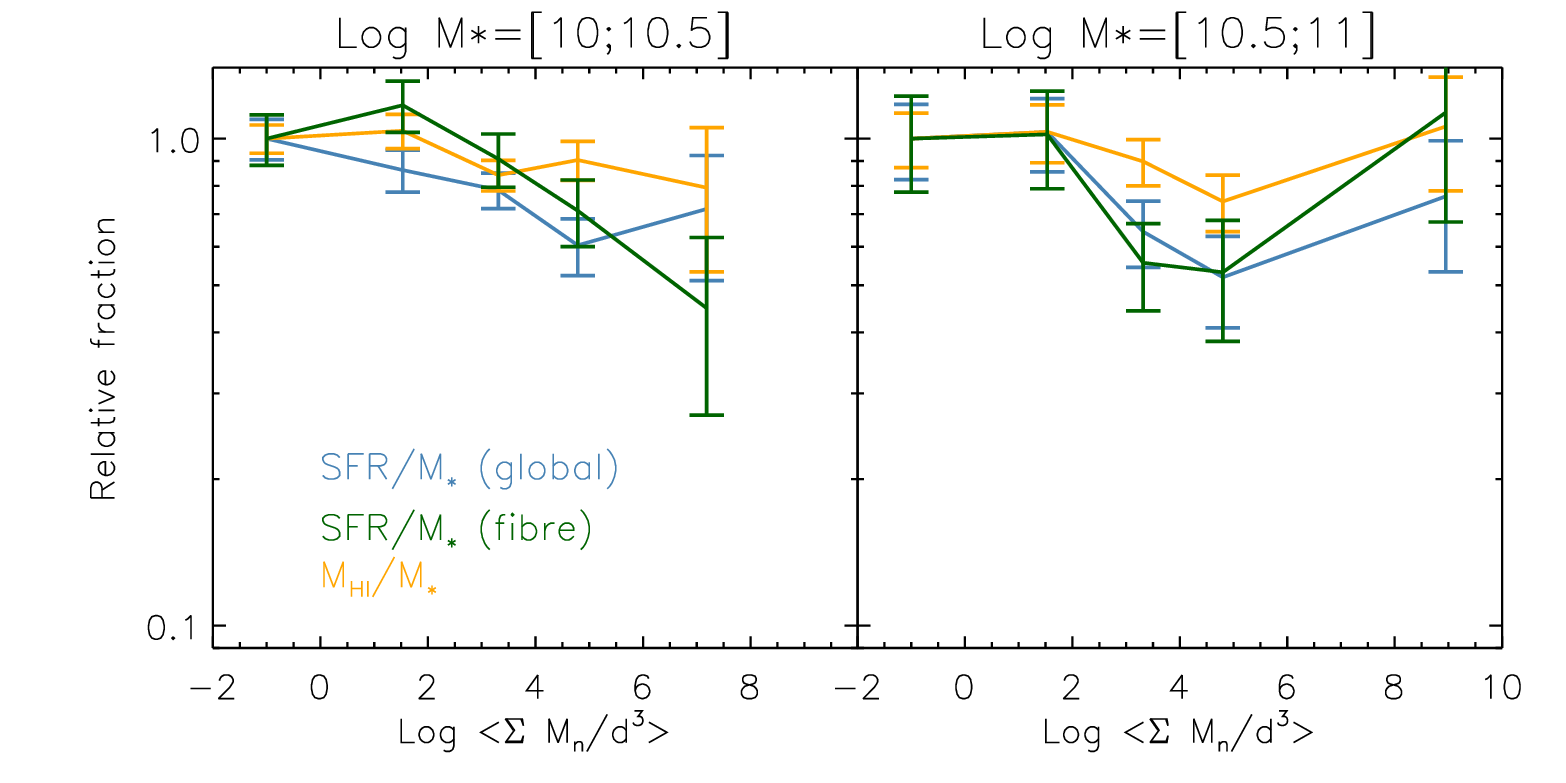}\\
    \includegraphics[width=8.5cm]{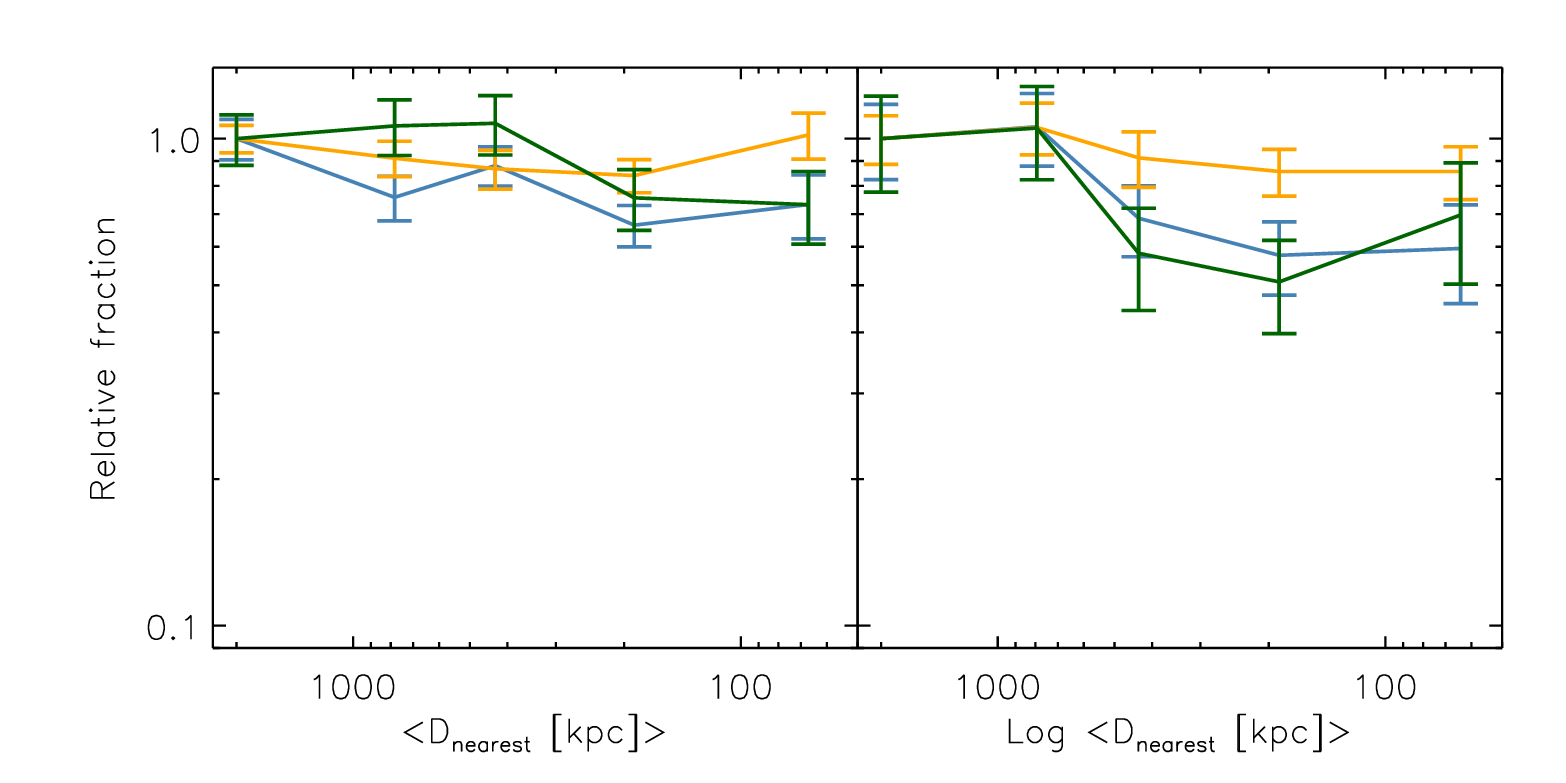}
  \end{tabular}\caption[Sky distribution of two groups 
    in our sample.]{Comparison of the relative dependence of the
        {\hi} gas fraction (orange), the global specific star
        formation rate (blue) and the fibre specific star formation
        rate (green) as a function of the summed  tidal force from the
        surrounding galaxies (top panel) and as a function of the
        distance to the nearest neighbour (bottom panel). The
        distances are 2-D projected ones. The left and right columns
        correspond to the 2 bins of {\Mst} as labelled at the top of
        the diagram. Error bars are evaluated by bootstrap re-sampling
        the galaxies in the stack.}\label{en:fig:fig09}
\end{figure}

One might ask whether tidal forces are likely to be  more effective
than ram-pressure  at stripping material from galaxies in lower
density environments. Tidal interactions between galaxies can affect
both the gas and the stars in these systems. The tidal force scales as
$M/d^3$, where $M$ is the mass of the neighbouring galaxy and $d$ is
its separation. In Figure \ref{en:fig:fig09} (top panel), we
  analyse trends in {\hi} gas mass fraction and specific star
  formation rate  as a  function of  the summed  tidal force from the
  surrounding galaxies. We use the projected distance as the measure
  of galaxy separation -- this is not strictly correct, but it is a
  better indicator of true separation than a 3-dimensional estimate  in rich groups and clusters where
galaxy peculiar velocities are large. We also look at trends in {\hi} gas fraction and sSFR as a function of the distance
to the nearest neighbour (Figure \ref{en:fig:fig09}, bottom panel).  
We find very weak effects as a function of both quantities. 
In addition, we do not observe the same ordering of {\hi} gas mass fraction, global
and fibre specific star formation rate seen in Figure 3 when we plot
these quantities as a function of the summed tidal force. 
However, we note that  the relevant tidal
force is not the current one, at the current separation, but the maximum
tidal force, experienced at the closest separation,
 which generally occurred in the past. Further investigation will be necessary
to make conclusive statements as to the role of tidal stripping in producing the
effects that we see.   

We also caution that the analysis presented in this work is only statistical in
nature. 
When stacking, it is  not possible to distinguish
between the effects of starvation or ram-pressure mechanisms in
individual galaxies.
Without  resolved gas maps, we have no information on the spatial extent or the morphology of
the atomic gas in the galaxies in our sample, which would more clearly diagnose  
ram-pressure stripping in individual systems.  
It would also be very interesting to study relative trends in
gas and stars not only as a function of local density, but as a function
of dark matter halo mass. This could be done by correlating the available ALFALFA 
data with group catalogues generated from the SDSS (Yang et al. 2007). 
With the full ALFALFA data set, 
it will be possible to measure the decrease in {\hi} for
increasing group/clustercentric distance, which would  put stronger
constraints on ram-pressure stripping mechanisms.     
Eventually, the all-sky surveys planned at the Westerbork telescope
(APERTIF; Verheijen et al. 2008) and at 
the Australian SKA Pathfinder telescope (ASKAP)  
will scan the sky in the
21 centimeter line with  much better sensitivity and resolution than
currently possible and produce datasets that are ideal for studying the environmentally-driven
processes that are important in understanding galaxy evolution.

\appendix
\newcommand{\appsection}[1]{\let\oldthesection\thesection
  \section{#1}\let\thesection\oldthesection}
\appsection{Correction for beam confusion}

\begin {figure*}
\centering
\includegraphics[width=14.cm]{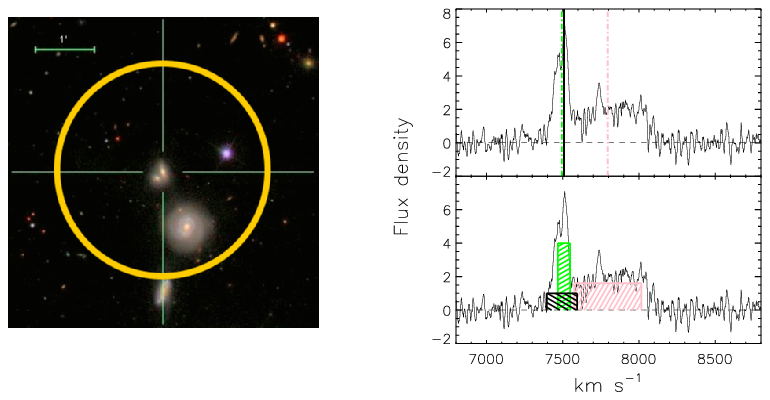}
\caption[Example of beam confusion]{
Example of possible signal confusion inside the Arecibo beam. {\em
Left:} SDSS image of the galaxy GASS 49727 and its companions.
The yellow circle indicates the 3.5 arcminute
Arecibo beam. {\em Right:} the spectrum obtained with
Arecibo. On top, the main target (black solid line) and the two
companions (coloured dotted lines) central velocities are flagged. 
On the  bottom, the shadowed regions
show how we would model the {\hi} signals, as
described in the text.}\label{fig:appb:eg} 
 \end{figure*}

The Arecibo telescope is a single dish of 305m in diameter and has a FWHM beam of  
$\sim$3.5 arcminutes at 21 cm, which corresponds 
to a physical scale of 150 kpc at the mean
redshift of the galaxies in our sample  ($z\simeq 0.037$).  
Confusion of signals coming from different galaxies within the beam at
similar redshift is thus of possible concern. 
An example of such a case is shown in Figure
\ref{fig:appb:eg}, where two additional companion galaxies are located within the beam.  
The SDSS image of the galaxy is
shown on the left and  the yellow circle indicates the Arecibo
beam size. In the right upper panel, the resulting
spectrum obtained with Arecibo is shown. The vertical lines flag the expected central
velocities of the three objects  from their SDSS redshifts. 

As discussed in Paper I, $\S$3, we visually inspected each spectrum and 
discarded the ones with a strong 
signal close to the galaxy but not centered at the expected
redshift, so the galaxy shown in \ref{fig:appb:eg} 
will actually be discarded from the sample. In our
analysis, confusion will arises from the stacking of non-detected {\hi} emission,  
or if the companion and main target have almost exactly the same redshift.

In previous work (Paper I, Fabello et al. 2011b)  we did not apply any
correction for possible confusion.  
For the environmental analysis, confusion
may be larger, especially in the high density bins where galaxies are
more clustered. In order to identify confused objects, we search
the MPA-JHU spectroscopic sample of galaxies with
{\Mst}$>3\times 10^9\,$M$_{\odot}$ for objects with projected distance
smaller than the beam FWHM and  velocity
separation smaller than 300 {\vel}. 
If the velocity difference is 
larger than this value, the {\hi}  signals will not overlap. Companions 
at large velocity separation may increase  the noise in the
baseline, but do not affect the measured gas content. Around 
20\% of \textit{sample A} targets have at least one   
companion that meet these criteria.  

In order to correct for confusion, we 
proceed as follows: 
\begin{itemize}
\item[(1)] We estimate the expected gas content of each companion,
  using the relation between colour, stellar mass surface density and {\hi}
  gas fraction derived by Zhang et al. (2009):
  \begin{eqnarray}
    \mathrm{Log}
    \left(\frac{\mathrm{M_{HI}}}{\mathrm{M_*}}\right)\,=\,-1.732\cdot
    (g-r)+0.215\cdot \mu_i-4.084,
  \end{eqnarray}
  where $\mu_i$ is the surface brightness in the $i$-band, and $g$ and
  $r$ are SDSS magnitudes corrected for Galactic extinction. 

\item[(2)] We estimate the actual signal contaminating   the stacked
  spectrum. First, we apply a correction factor to the emission from the companion
  ($f_1$) using  the projected  distance between the
  target and the companion. The beam profile can be 
  approximated with  a 2D Gaussian with $\sigma_x=(2\sqrt{2\cdot
    \mathrm{ln}2})^{-1}\times3.3$arcminutes and $\sigma_y=(2\sqrt{2\cdot
    \mathrm{ln}2})^{-1}\times3.8$arcminutes, so that its response decreases at the
  edges. The bigger companion in Figure \ref{fig:appb:eg}, for
    example, lies at a projected distances of $x\simeq0.4\,$arcminutes, and
  $y\simeq1\,$arcminutes from the target. Therefore, 
  $f_1=\exp[-0.5\cdot(x/\sigma_x)^2-0.5\cdot(y/\sigma_y)^2]$=0.8 of its
  flux would be  recorded. 

  Likewise, only part of the signal from the companions will  actually 
  overlap with the main target in  velocity space. To estimate this second
  correction factor ($f_2$), we calculate the expected
  {\hi} line widths of both the main target ($w_t$) and the companion
  ($w_c$),  assuming a box-shape profile.  To
  evaluate the observed width we use a Tully-Fisher relation as in Paper I
  ($\S\,$3.2), where:  $w_{obs}=w_{TF}\cdot \sin(incl)$, and $w_{TF}$
  is evaluated following the relation from Giovanelli et al. (1997), and using the SDSS 
  $i$-band magnitude, $k$-corrected and corrected for Galactic and
  internal extinction (as in equations 11 and 12 in Giovanelli et al. 1997). 
  The correction factor $f_2$           
   is given by the velocity overlap ($\Delta w$) between
  $w_t$ and $w_c$: $f_2=\Delta w/w_c$.  
  As an example, in Figure 
  \ref{fig:appb:eg} (bottom spectrum) the dashed regions represent how
  we would have modeled the  three signals contributing to the
  spectrum. In the example, the entire flux from the green companion 
  contributes  to the measured signal ($f_2=1$), because it 
  overlaps fully with the  main target emission (black region). In contrast, 
  the pink companion contributes only  a very small
  fraction of its emission to the signal  (overlap of $f_2=0.06$). 

\item[(3)]  For companions with separation from the main target $\Delta v > 50\,${\vel},
    we check that the {\hi} mass predicted using Eq. A1 lies below the ALFALFA upper limit.
    If above, such companions should have been flagged during visual
    inspection of the spectrum. Failure to detect such a companion actually implies that
    the Zhang et al. estimate of the {\hi} content is too high;
    we then reset our estimate of the gas mass of the companion
    to the actual ALFALFA upper limit.

\item[(4)] Finally, we subtract the contributions from all the confused
companions to the {\hi} mass measured from the stacked spectrum, as follows.
  
  Because of the weight we apply to the spectra of the individual
  galaxies, a gas fraction measured from the stacked spectrum is  (Paper I, $\S\,$3.3):    
\begin{equation}\label{gf}
  \frac{\mathrm{M_{HI}}}{\mathrm{M_*}}=\frac{2.356\times10^5}{\Sigma_i\mathrm{w}_i}\Sigma_i\frac{\mathrm{D^2_L}(z_i)}{(1+z_i)}\frac{S_i}{\mathrm{M}_{*;i}}\mathrm{w}_i, 
\end{equation} 
 where $D_L(z)$ is the luminosity distance, $S$ the integrated
{\hi} flux and $w=1/rms^2$, and $i$ the index running over the
individual galaxies. The total signal $S_i$ is
actually the real emission from the 
main target ($S_t$) plus the ones from the confused objects ($S_c$),
weighted for the two factors described in point (2): 
\begin{equation*}
  S_i\,=\,S_t+\Sigma_c \,f_{1;c}f_{2;c}S_c.
\end{equation*} 
We can rewrite equation A2  as
\begin{equation*}
  \frac{\mathrm{M_{HI}}}{\mathrm{M_*}}= \left(\frac{\mathrm{M_{HI}}}{\mathrm{M_*}}\right)_t+ \frac{2.356\times10^5}{\Sigma_i\mathrm{w}_i}\Sigma_i\frac{\mathrm{D^2_L}(z_i)}{(1+z_i)}\frac{\Sigma_c \,f_{1;c}f_{2;c}S_c}{\mathrm{M}_{*;i}}\mathrm{w}_i. 
\end{equation*} 
And finally, if we substitute the companions' gas fractions estimated from
photometry (gf$_c$), as described in equation A1 , we obtain:
\begin{eqnarray*}%
&
  \left(\frac{\mathrm{M_{HI}}}{\mathrm{M_*}}\right)_t=\frac{\mathrm{M_{HI}}}{\mathrm{M_*}}+
  \frac{1}{\Sigma_i\mathrm{w}_i}\Sigma_i\left[\frac{\mathrm{D^2_L}(z_i)}{(1+z_i)}\frac{\mathrm{w}_i}{\mathrm{M}_{*;i}}\Sigma_c
    \left(\frac{f_{1;c}f_{2;c}\mathrm{gf}_c 
        \mathrm{M}_{*;c}(1+z_c)}{\mathrm{D^2_L}(z_c)}\right)\right] 
\end{eqnarray*} 
\end{itemize}

\normalsize
As mentioned in the paper,
confusion corrections are always small. Even in the highest density bins, 
the correction factor is smaller than few percent.

\section*{Acknowledgments}
We thank the referee of this paper, Jeff Kenney, for his helpful and
constructive comments. 

SF wishes to thank Luca Cortese for suggestions and useful
discussions. 

We thank the many members of the ALFALFA team who have contributed to
the acquisition and processing of the ALFALFA dataset over the last
six years. 

CL is supported by NSFC  (no.   11173045),  Shanghai Pujiang
Programme (no. 11PJ1411600), the CAS/SAFEA International
Partnership Program for  Creative  Research  Teams  (KJCX2-YW-T23),
the  100 Talents  Program of  Chinese Academy  of Sciences
(CAS) and the exchange program between Max Planck Society
and CAS.

RG and MPH are supported by NSF grant AST-0607007 and by a grant from
the Brinson Foundation.

The Arecibo Observatory is operated by SRI International under a
cooperative agreement with the National Science Foundation
(AST-1100968), and in alliance with Ana G. Mandez-Universidad
Metropolitana, and the Universities Space Research Association.


\end{document}